\documentclass[reviewcopy]{elsart}
\usepackage{amsmath,amssymb}
\usepackage{graphicx,epsfig}

\begin{document}

\begin{frontmatter}

\title{Thermodynamic properties of a classical $d$-dimensional 
spin-$S$ Heisenberg ferromagnet with long-range interactions
via the spectral density method}

\author[unisa,infm]{A. Cavallo},
\author[unisa,infm]{F. Cosenza \corauthref{cosfab}} and
\ead{cosfab@sa.infn.it}
\author[unisa,infm]{L. De Cesare}
\corauth[cosfab]{}

\address[unisa]{Dipartimento di Fisica "E.R. Caianiello", Universit\`{a} degli Studi di Salerno,\\ 
Via S. Allende, 84081 Baronissi (SA), Italy}
\address[infm]{Istituto Nazionale di Fisica della Materia (INFM), Unit\`{a} di Salerno, Italy}

\begin{abstract}
The thermodynamic properties of a classical $d$-dimensional 
spin-$S$ Heisenberg ferromagnet, with long-range interactions decaying 
as $r^{-p}$ and in the presence of an external magnetic field, 
is investigated by means of the spectral density method in the framework 
of classical statistical mechanics. 
We find that long-range order exists at finite temperature for $d<p<2d$ 
with $d\leq 2$ and for $p>d$ with $d>2$, consistently with known theorems.
Besides, the related critical temperature is determined and a study of the 
critical properties is performed.
\end{abstract}


\end{frontmatter}


\section{Introduction}
The quantum two-time Green function technique has provided a very powerful
tool in condensed matter physics for exploring the equilibrium and transport
properties of a wide variety of many-body systems. Within this framework,
the equation of motion method (EMM) and the spectral density method (SDM)
allow to obtain reliable approximations to treat typically unperturbative
problems \cite{Zubarev60,Tyablikov67,Kalashnikov73}. The pioneering
introduction of the two-time Green functions and the EMM in classical
statistical mechanics by Bogoljubov and Sadovnikov \cite{Bogoljubov62}, has
opened the concrete possibility to describe classical and quantum systems on
the same footing. Nextly, a classical version of the SDM (CSDM) has been
also formulated in extensive \cite{Caramico81,Campana83,Campana84} and
nonextensive \cite{Cavallo01} classical statistical mechanics and applied to
classical magnetic chains \cite{Caramico81,Campana83,Campana84,Cavallo01}.
The last method, which seems to present many advantages with respect to the
most conventional EMM 
\cite{Kalashnikov73,Caramico81,Campana83,Campana84,Cavallo01}, offers a robust
instrument for systematic and well tested approximations \cite{Kalashnikov73}
to explore the macroscopic properties of classical many-body systems as
well. Unfortunately, both the mentioned methods have not received the due
consideration in the classical context and further developments and
applications are desiderable. Along this direction, in this paper we apply
the CSDM to investigate the thermodynamic properties of a highly nontrivial $%
d$-dimensional classical spin-$S$ Heisenberg ferromagnet with long-range
interactions decaying as $r^{-p}$ ($p>d$) with the distance $r$ between
spins in the presence of an external magnetic field. The choice of this
model, besides its intrinsic theoretical interest, is motivated also by its
relevance in the description of many materials \cite
{Steiner76,Reither80,Sepliarky01} of experimental and technological
importance and, in particular, in the understanding of the critical behavior
of magnetic systems when long-range exchange interactions are believed to
play an important role. In general, realistic microscopic magnetic models
with long-range interactions are difficult to be studied in a reliable way.
Since the exact solution for the spin-$1/2$ quantum Heisenberg chain with an
inverse-square exchange $(p=2)$ and without an external field was found
independently by Haldane \cite{Haldane88} and Shastry \cite{Shastry88}, only
a limited lot of information has been acquired about the low-temperature
properties, the existence of long-range order (LRO) and the related critical
behavior for different values of the decaying exponent $p$ and the
dimensionality $d$.

Recently, the critical properties at finite temperature of $d$-dimensional
quantum Heisenberg models, with interactions of the type here considered
(here often named ''long-range spin models'') have been studied using
microscopic techniques \cite{Nakano94a,Nakano94b,Nakano95,Bruno01} and Monte
Carlo simulations \cite{Vassiliev01}. A summary of the known features sounds
as follows. The one- and two-dimensional long-range quantum spin-$1/2$
Heisenberg ferromagnets in absence of an external magnetic field were
investigated by Nakano and Takahashi using the so called modified spin-wave
theory \cite{Nakano94a} and the Schwinger-boson mean-field approximation 
\cite{Nakano94b}. Further information were derived for the $d$-dimensional
case by means of the EMM for the two-time Green functions using the
Tyablikov decoupling procedure \cite{Nakano95}. Monte Carlo simulations for
the two-dimensional quantum spin-$1/2$ Heisenberg model have been also
performed for $2<p\leq 6$ \cite{Vassiliev01}. This scenario, has been
recently enriched by an extension \cite{Bruno01} of the Mermin-Wagner
theorem \cite{MerminWagner66} for the existence of ferromagnetic (FM) LRO at
finite temperature in quantum Heisenberg and XY models in $d(=1,2)$
dimensions with $r^{-p}$- and oscillatory- interactions.

Classical long-range spin-s Heisenberg FM models have attracted great
attention, too. It has been proved that LRO exist in $d(=1,2)$ dimension
when $d<p<2d$ \cite{Kunz76,Frolich78} and is destroyed at all finite
temperatures for $p\geq 2d$ \cite{Rogers81,Simon81,Pfister81}. Similar
results were obtained for the spherical model \cite{Joyce66} and the present
scenario of the critical properties is largely based on renormalization
group calculations for the classical $n$-vector model \cite
{Fisher72,Kosterlitz76}. Classical long-range antiferromagnetic (AFM) models
have been studied less extensively. The available rigorous results \cite
{Kunz76,Frolich78,Rogers81} suggest orientational disorder at all finite
temperatures when $p\geq 2d$, but no theorem exists entailing existence or
absence of LRO for $d<p<2d$. Monte Carlo simulations have been also
performed for both $d(=1,2)$-dimensional classical FM (for $p=2d$ \cite
{Romano89}) and AFM (for $p=3/2$ and $p=3$ with $d=1$ and $d=2$ respectively 
\cite{Romano90}) Heisenberg long-range models. The results confirm that
FM-LRO survives at finite temperature provided $d<p<2d$ and allow to
conjecture that no AFM-LRO exists at all finite temperatures for $p>d$.
Spin-wave studies \cite{Romano90} agree with last conjecture but no
definitive statement can be drawn at the present stage. In a quite recent
work \cite{Cavallo02} the thermodynamics and the critical properties of the
classical long-range spin-$S$ Heisenberg FM chain in the presence of an
external field have been systematically studied using the CSDM to lowest
order of approximation. The results are in good agreement with previous
available analytical and numerical investigations. As mentioned before, here
we use the same method to explore a $d$-dimensional Heisenberg FM model. As
we shall see, a rich phase diagram appears where all the above mentioned
scenario is reproduced and extended.

The paper is organized as follows. In Sec. 2 we introduce the model and the
moment equations (ME's) for the spectral density (SD) to lowest order of
approximation. The main low-temperature properties are studied in Sec. 3 and
the existence of LRO in different regions of the $(d-p)$-plane is also shown
in a transparent way. Sec. 4 is devoted to the critical properties and the
low-temperature paramagnetic susceptibility of the model. Finally, in Sec.
5, some concluding remarks are drawn.

\section{The model and the moment equations for the spectral density}
A classical $d$-dimensional spin-$S$ Heisenberg ferromagnet with long-range
interactions is described by the Hamiltonian: 
\begin{equation}
H=-\frac{1}{2}\sum_{i,j=1}^{N}J_{ij}\mathbf{S}_{i}\cdot \mathbf{\ S}%
_{j}-h\sum_{i=1}^{N}S_{i}^{z}.  \label{eq:1model}
\end{equation}
Here, $N$ is the number of sites of the hypercubic lattice with unitary
spacing, $\left\{ \mathbf{S}_{i};i=1,...,N\right\} $ are the classical
spins, $h$ is the external magnetic field and the spin-spin interaction, in
view of the thermodynamic limit as $N\rightarrow \infty $, is assumed to be $%
J_{ij}=J/r_{ij}^{p}$, where $r_{ij}\equiv |\mathbf{r}_{i}-\mathbf{r}_{j}|$
and $J>0$ measures the strength of the coupling. The extreme case $%
p\rightarrow \infty $ corresponds to the standard nearest-neighbor
interaction while the mean field approximation is obtained when $p=0$
(replacing $J$ by $J/N$). For this type of interaction, the thermodynamical
limit $N\rightarrow \infty $ is well defined only for $p>d$, while for $%
p\leq d$ the ground state of the system has an infinite energy per particle
as $N\rightarrow \infty $ and the conventional statistical mechanics cannot
be directly applied.

The classical spins model (\ref{eq:1model}) can be appropriately described
by the set of $2N$ canonical variables $\left\{ \varphi
_{j},S_{j}^{z}\right\} $ where $\varphi _{j}$ is the angle between the
projection of the spin vector $\mathbf{S}_{j}$ in the ($x-y$)-plane and the $%
x$ axis. For practical calculations we find convenient to introduce the new
variables $S_{j}^{\pm }=S_{j}^{x}\pm iS_{j}^{y}$, so that $%
S_{j}^{2}=(S_{j}^{z})^{2}+S_{j}^{+}S_{j}^{-}$. Then, the Hamiltonian %
\eqref{eq:1model} can be rewritten as 
\begin{equation}
H=-\frac{1}{2N}\sum_{\mathbf{k}}J\left( \mathbf{k}\right) \left( S_{\mathbf{k%
}}^{+}S_{-\mathbf{k}}^{-}+S_{\mathbf{k}}^{z}S_{-\mathbf{k}}^{z}\right)
-hS_{0}^{z}  \label{eq:2model}
\end{equation}
involving the Fourier components of the spins and the exchange interaction
defined by: 
\begin{equation}
\mathbf{S}_{\mathbf{k}}=\sum_{j=1}^{N}e^{-i\mathbf{k}\cdot \mathbf{r}_{j}}%
\mathbf{S}_{j},\text{ \ }J\left( \mathbf{k}\right) =\sum_{j=1}^{N}e^{i%
\mathbf{k}\cdot (\mathbf{r}_{i}-\mathbf{r}_{j})}J_{ij}\text{,}
\label{eq:3FourierTr}
\end{equation}
where $\mathbf{k}$ denotes a wave vector in the $d$-dimensional Fourier
space. The sum in Eq. (\ref{eq:2model}) is restricted to the first Brillouen
zone ($1BZ$) of the lattice. The Poisson brackets for the spin Fourier
components relevant for us, are: 
\begin{equation}
\left\{ S_{\mathbf{k}}^{\pm },S_{\mathbf{k}^{\prime }}^{z}\right\} =\pm iS_{%
\mathbf{k}+\mathbf{k}^{\prime }}^{\pm },\text{ \ }\left\{ S_{\mathbf{k}%
}^{+},S_{\mathbf{k}^{\prime }}^{-}\right\} =-2iS_{\mathbf{k}+\mathbf{k}%
^{\prime }}^{z}\text{ .}
\end{equation}

In the context of the CSDM \cite
{Caramico81,Campana83,Campana84,Cavallo01,Cavallo02}, all the
thermodynamical properties of the model can be derived using the SD defined
by: 
\begin{equation}
\Lambda _{\mathbf{k}}(\omega )=-i\left\langle \left\{ S_{-\mathbf{k}}^{-},S_{%
\mathbf{k}}^{+}\left( t\right) \right\} \right\rangle _{\omega
}=-i\int_{-\infty }^{\infty }dte^{i\omega t}\left\langle \left\{ S_{-\mathbf{%
k}}^{-},S_{\mathbf{k}}^{+}\left( t\right) \right\} \right\rangle \text{,}
\end{equation}
where the classical dynamical variables depend on time through the conjugate
canonical coordinates $\left\{ \varphi _{j}\left( t\right) ,S_{j}^{z}\left(
t\right) ;j=1,...N\right\} $. Here $A\left( t\right) =e^{iLt}A\left(
0\right) $, $L=i\left\{ H,...\right\} $ is the Liouville operator and $%
\left\langle ...\right\rangle $ denotes a canonical ensemble average.
According to the spirit of the CSDM, we try to determine it to the lowest
order in the form \cite{Campana84,Cavallo02}: 
\begin{equation}
\Lambda _{\mathbf{k}}(\omega )=2\pi \lambda _{\mathbf{k}}\delta (\omega
-\omega _{\mathbf{k}})\text{,}  \label{eq:SD_delta}
\end{equation}
where the unknown parameters $\lambda _{\mathbf{k}}$ and $\omega _{\mathbf{k}%
}$ are to be calculated solving the first two moment equations (ME's): 
\begin{align}
\int_{-\infty }^{+\infty }\frac{d\omega }{2\pi }\Lambda _{\mathbf{k}}(\omega
)& =-i\left\langle \left\{ S_{-\mathbf{k}}^{-},S_{\mathbf{k}}^{+}\right\}
\right\rangle =2Nm,  \label{eq:ME1} \\
\int_{-\infty }^{+\infty }\frac{d\omega }{2\pi }\omega \Lambda _{\mathbf{k}%
}\left( \omega \right) & =-i\left\langle \left\{ S_{-\mathbf{k}}^{-},\left\{
H,S_{\mathbf{k}}^{+}\right\} \right\} \right\rangle   \nonumber \\
& =\frac{1}{N}\sum_{\mathbf{k}^{\prime }}(J(\mathbf{k}^{\prime })-J(\mathbf{k%
}-\mathbf{k}^{\prime }))(\left\langle S_{\mathbf{k}^{\prime }}^{+}S_{-%
\mathbf{k}^{\prime }}^{-}\right\rangle +2\left\langle S_{\mathbf{k}^{\prime
}}^{z}S_{-\mathbf{k}^{\prime }}^{z}\right\rangle )+2Nmh  \label{eq:ME2}
\end{align}
with $m=\left\langle S_{j}^{z}\right\rangle $ the magnetization per spin.

To close this system, we should express all the unknown quantities in terms
of the SD. One can show\cite{Campana84,Cavallo02} that the transverse
correlation function $\langle S_{\mathbf{k}}^{+}S_{-\mathbf{k}}^{-}\rangle $%
, appearing on the right--hand side of Eq. (\ref{eq:ME2}), can be exactly
expressed in terms of the SD by the relation: 
\begin{equation}
\langle S_{\mathbf{k}}^{+}S_{-\mathbf{k}}^{-}\rangle =T\int_{-\infty
}^{\infty }\frac{d\omega }{2\pi }\frac{\Lambda _{\mathbf{k}}(\omega )}{%
\omega }.  \label{eq:S+S-_SD}
\end{equation}
Then, taking into account the one-$\delta $ ansatz (\ref{eq:SD_delta}), we
have 
\begin{equation}
\left\langle S_{\mathbf{k}}^{+}S_{-\mathbf{k}}^{-}\right\rangle =\frac{2NmT}{%
\omega _{\mathbf{k}}}.  \label{eq:S+S-}
\end{equation}
On the contrary, the longitudinal correlation function $\left\langle S_{%
\mathbf{k}}^{z}S_{-\mathbf{k}}^{z}\right\rangle $ cannot be exactly related
to the spectral density $\Lambda _{\mathbf{k}}(\omega )$. So, to close the
system of ME's (\ref{eq:ME1}),(\ref{eq:ME2}) we should solve another moment
problem by introducing a new SD. The simplest way to avoid this difficulty
is to resort to the decoupling $\left\langle S_{\mathbf{k}}^{z}S_{-\mathbf{k}%
}^{z}\right\rangle \approx \left\langle S_{\mathbf{k}}^{z}\left\rangle \cdot
\right\langle S_{-\mathbf{k}}^{z}\right\rangle =N^{2}m^{2}\delta _{\mathbf{k}%
,0}$, which means to neglect the correlations between the Fourier components
of the $S_{j}^{z}$. Of course, the approximation is appropriate to describe
thermodynamic regimes with finite magnetization as under near saturation
conditions. In Sec. 4 we will introduce a different procedure which allows
us to describe also, in a reliable way, regimes with zero or near zero
magnetization. As final step one must express the magnetization $m$ in terms
of the $\Lambda _{\mathbf{k}}(\omega )$. This is not a simple problem for a
classical spin system, also when $S=1/2$ (for which the exact relation $%
S_{j}^{z}=1/2-S_{j}^{+}S_{j}^{-}$ exists in the quantum counterpart).
However, as shown in Refs. \cite{Campana84,Cavallo02} within the spirit of
the SDM, one can use the expression: 
\begin{equation}
m^{2}=\frac{S^{2}-\frac{3}{2N^{2}}\sum_{\mathbf{k}}\left\langle S_{\mathbf{k}%
}^{+}S_{\mathbf{-k}}^{-}\right\rangle }{1-\frac{1}{2S^{2}N^{2}}\sum_{\mathbf{%
k}}\left\langle S_{\mathbf{k}}^{+}S_{-\mathbf{k}}^{-}\right\rangle }\text{,}
\label{eq:m2}
\end{equation}
which is appropriate for all thermodynamic regimes and reduces, correctly,
to the near saturation relation $m\simeq S-\left\langle
S_{j}^{+}S_{j}^{-}\right\rangle /(2S)=S-\sum_{\mathbf{k}}\left\langle S_{%
\mathbf{k}}^{+}S_{-\mathbf{k}}^{-}\right\rangle /(2SN^{2})$ arising from the
identity $S^{2}=\left( S_{j}^{z}\right) ^{2}+S_{j}^{+}S_{j}^{-}$ with $%
\left| S_{j}^{+}S_{j}^{-}\right| /S^{2}\ll 1$. Then, from Eq. (\ref{eq:S+S-}%
), as a consequence of Eq. (\ref{eq:S+S-_SD}) and the ansatz (\ref
{eq:SD_delta}), Eq. (\ref{eq:m2}) yields: 
\begin{equation}
m^{2}=\frac{S^{2}-\frac{3Tm}{N}\sum_{\mathbf{k}}\frac{1}{\omega _{\mathbf{k}}%
}}{1-\frac{Tm}{NS^{2}}\sum_{\mathbf{k}}\frac{1}{\omega _{\mathbf{k}}}}\text{.%
}
\end{equation}

With the above ingredients, Eqs. (\ref{eq:ME1})-(\ref{eq:ME2}) become a
closed system to be solved self-consistently. By introducing conveniently
the dimensionless variables $\sigma =m/S$, $\overline{T}=T/JS^{2}$, $%
\overline{h}=h/JS$ and $\overline{\omega }_{k}=\omega _{k}/JS$, the ME's
reduce to: 
\begin{align}
\overline{\omega }_{\mathbf{k}}& =\overline{h}+\sigma \Omega ^{(p)}(\mathbf{k%
})+\frac{\overline{T}}{N}\sum_{\mathbf{k}^{\prime}}\frac{\Omega ^{(p)}(%
\mathbf{k}-\mathbf{k}^{\prime })-\Omega ^{(p)}(\mathbf{k}^{\prime })}{%
\overline{\omega }_{\mathbf{k}^{\prime }}},  \label{eq:se1} \\
\sigma ^{2}& =\frac{1-3\sigma \frac{\overline{T}}{N}\sum_{\mathbf{k}}\frac{1%
}{\omega _{\mathbf{k}}}}{1-\sigma \frac{\overline{T}}{N}\sum_{\mathbf{k}}%
\frac{1}{\omega _{\mathbf{k}}}},  \label{eq:sigma1}
\end{align}
with 
\begin{equation}
\Omega ^{(p)}(\mathbf{k})=\sum_{\mathbf{r}}\frac{1-\cos \mathbf{k}\cdot 
\mathbf{r}}{|\mathbf{r}|^{p}}.  \label{eq:Omega(k)}
\end{equation}
In the thermodynamic limit $N\rightarrow \infty $ one must replace the sum $%
\sum_{\mathbf{k}}(...)/N$ by an integral $\int_{1BZ}(...)d^{d}k/(2\pi )^{d}$
in the $d$-dimensional $\mathbf{k}$-space and hence Eqs.(\ref{eq:se1}) and (%
\ref{eq:sigma1}) can be rewritten as: 
\begin{align}
\overline{\omega }_{\mathbf{k}}& =\overline{h}+\sigma \Omega ^{(p)}(\mathbf{k%
})R(\mathbf{k})\text{,}  \label{eq:se_R} \\
\sigma ^{2}& =\frac{1-3\overline{T}\sigma \int_{1BZ}\frac{d^{d}k}{(2\pi )^{d}%
}\frac{1}{\overline{\omega }_{\mathbf{k}}}}{1-\overline{T}\sigma \int_{1BZ}%
\frac{d^{d}k}{(2\pi )^{d}}\frac{1}{\overline{\omega }_{\mathbf{k}}}}\text{.}
\label{eq:sigma}
\end{align}
with 
\begin{equation}
R(\mathbf{k})=1+\frac{\overline{T}}{\sigma }\int_{1BZ}\frac{d^{d}k^{\prime }%
}{(2\pi )^{d}}\frac{\Omega ^{(p)}(\mathbf{k}-\mathbf{k}^{\prime })-\Omega
^{(p)}(\mathbf{k}^{\prime })}{\overline{\omega }_{\mathbf{k}^{\prime
}}\Omega ^{(p)}(\mathbf{k})}\text{.}  \label{eq:R(k)}
\end{equation}
The solution of the problem is complicated and one must consider asymptotic
regimes for obtaining explicit results or use numerical calculations.
Moreover, we can note that these equations have physical meaning only if $%
\sigma \neq 0$ \cite{Cavallo02}. Hence, they cannot describe the critical
behavior of the system or a paramagnetic phase in zero external field. As we
will see, in Sec. 3 this problem can be simply overcome with an appropriate
modification of Eq. (\ref{eq:R(k)}).

\section{Low--temperature properties}
We first examine analytically the low--temperature solution of Eqs. (\ref
{eq:se_R})-(\ref{eq:R(k)}). The expressions of $\sigma $ and $\overline{%
\omega }_{\mathbf{k}}$ to the first order in the reduced temperature $%
\overline{T}$, which allow us to capture some relevant aspects of the
low-temperature physics of the model (\ref{eq:1model}), are given by: 
\begin{align}
\overline{\omega }_{\mathbf{k}}& \simeq \overline{h}+\Omega ^{(p)}(\mathbf{k}%
)-\overline{T}\left\{ I_{d}^{(p)}(\overline{h})\Omega ^{(p)}(\mathbf{k})+%
\mathcal{I}_{d}^{(p)}(\overline{h},\mathbf{k})\right\} ,  \label{eq:omg_lowT}
\\
\sigma & \simeq 1-\overline{T}I_{d}^{(p)}(\overline{h}),
\label{eq:sigma_lowT}
\end{align}
where 
\begin{eqnarray}
I_{d}^{(p)}(\overline{h}) &=&\int_{1BZ}\frac{d^{d}k}{(2\pi )^{d}}\frac{1}{%
\overline{h}+\Omega ^{(p)}(\mathbf{k})},  \label{eq:funcI(h)} \\
\mathcal{I}_{d}^{(p)}(\overline{h},\mathbf{k}) &=&\int_{1BZ}\frac{%
d^{d}k^{\prime }}{(2\pi )^{d}}\frac{\Omega ^{(p)}(\mathbf{k}^{\prime
})-\Omega ^{(p)}(\mathbf{k}-\mathbf{k}^{\prime })}{\overline{h}+\Omega
^{(p)}(\mathbf{k}^{\prime })}.
\end{eqnarray}
At first, we assume $\overline{h}\neq 0$ so that no convergency problem for
the above integrals in the $\mathbf{k}$-space occurs. To calculate
analytically these integrals, one needs an explicit expression of the
function $\Omega ^{(p)}(\mathbf{k})$ which, unfortunately, cannot be
obtained in terms of elementary functions in the whole $1BZ$, for arbitrary
values of the parameter $p>d$. Nevertheless, for sufficiently small values
of the external magnetic field, the dominant contribution to the integrals
in Eqs.(\ref{eq:omg_lowT}) and (\ref{eq:sigma_lowT}) arises from the low
wave-vector excitations as close to the critical temperature (see Sec. 4).
Then, we can obtain an explicit estimate of $\sigma $ and $\overline{\omega }%
_{\mathbf{k}}$ assuming the dominant behavior of $\Omega ^{(p)}(\mathbf{k})$
in the $1BZ$ as $\mathbf{k}\rightarrow 0$, provided that the coefficients of 
$\overline{T}$ in Eqs. (\ref{eq:omg_lowT}) and (\ref{eq:sigma_lowT}) remain
finite. Keeping this in mind, one can show \cite
{Nakano95,Bruno01,Romano89,Romano90} that for $p>d$ we have for $\Omega
^{(p)}(\mathbf{k})$ the low-$k$ expansions: 
\begin{equation}
\Omega ^{(p)}(\mathbf{k})\simeq \left\{ 
\begin{array}{ll}
A_{d}k^{p-d}+B_{d}k^{2}+O(k^{4}),\text{ } & p\neq d+2 \\ 
C_{d}k^{2}\ln (\Lambda /k)+O(k^{4}), & p=d+2.
\end{array}
\right.  \label{eq:Omega_lowk}
\end{equation}
The explicit expressions of the coefficients $A_{d}$, $B_{d}$ and $C_{d}$,
which depend in a cumbersome way on the dimensionality $d$, the exponent $p$
and a wave-vector cut-off $\Lambda $ related to the geometrical definition
of $1BZ$, are inessential at this stage and will be omitted. However, for
case $d=2$, they will be explicitly given in Sec. 4 where a comparison with
some analytical and Monte Carlo predictions is performed.

Taking into account Eq. (\ref{eq:Omega_lowk}), the integral $I_{d}^{(p)}(%
\overline{h})$ can be explicitely estimated and we have the following
low-temperature representation for the reduced magnetization 
\begin{equation}
\sigma \simeq 1-\overline{T}\left\{ 
\begin{array}{ll}
\frac{1}{\overline{h}}\text{ }_{2}F_{1}\left( 1,\frac{d}{p-d},\frac{p}{p-d};-%
\frac{A_{d}\Lambda ^{p-d}}{\overline{h}}\right) , & d<p<d+2 \\ 
K_{d}\int_{0}^{\Lambda }dkk^{d-1}\left[ \overline{h}+C_{d}k^{2}\ln (\Lambda
/k)\right] ^{-1}, & p=d+2 \\ 
\frac{1}{\overline{h}}\text{ }_{2}F_{1}\left( 1,\frac{d}{2},1+\frac{d}{2};-%
\frac{B_{d}\Lambda ^{2}}{\overline{h}}\right) , & p>d+2
\end{array}
\right.  \label{eq:sigma_explicit}
\end{equation}
where$\text{ }_{2}F_{1}\left( a,b,c;z\right) $ is the hypergeometric
function, $K_{d}=2^{1-d}\pi ^{-d/2}/\Gamma \left( d/2\right) $ and $\Gamma
\left( z\right) $ is the gamma function. An analogous calculation of $%
\mathcal{I}_{d}^{(p)}(\overline{h},\mathbf{k})$ in the expression (\ref
{eq:omg_lowT}) for $\overline{\omega }_{\mathbf{k}}$ is rather complicated.
However, an explicit estimate of $\overline{\omega }_{\mathbf{k}}$ based on
the expansions (\ref{eq:Omega_lowk}) is irrelevant for next developments.
The low-temperature susceptibility $\chi $ can be now easily obtained from
Eq. (\ref{eq:sigma_explicit}) by derivation with respect to the reduced
magnetic field $\overline{h}$. For the reduced susceptibility $\overline{%
\chi }=\chi /J$ we have: 
\begin{equation}
\overline{\chi }=\left. \frac{\partial \sigma }{\partial \overline{h}}%
\right| _{\overline{T}}\simeq \overline{T}\times \left\{ 
\begin{array}{ll}
\frac{1}{\overline{h}^{2}}\text{ }_{2}F_{1}\left( 2,\frac{d}{p-d},\frac{p}{%
p-d},-\frac{A_{d}\Lambda ^{p-d}}{\overline{h}}\right) , & d<p<d+2 \\ 
K_{d}\int_{0}^{\Lambda }dkk^{d-1}\left[ \overline{h}+C_{d}k^{2}\ln (\Lambda
/k)\right] ^{-2}, & p=d+2 \\ 
\frac{1}{\overline{h}^{2}}\text{ }_{2}F_{1}\left( 2,\frac{d}{2},1+\frac{d}{2}%
,-\frac{B_{d}\Lambda ^{2}}{\overline{h}}\right) , & p>d+2.
\end{array}
\right.   \label{eq:Chi}
\end{equation}

The above low-$T$ expressions have a physical meaning for $\overline{T}$ and 
$\overline{h}$ in the near saturation regime ($\sigma \simeq 1$) and also
for $\overline{h}\rightarrow 0$ when long--range order occurs. As mentioned
before, the integrals in Eqs. (\ref{eq:omg_lowT}) and (\ref{eq:sigma_lowT}),
and hence the functions in Eqs. (\ref{eq:sigma_explicit})-(\ref{eq:Chi}),
could diverge in the limit $\overline{h}\rightarrow 0$ for particular values
of the exponent $\alpha $ and the dimensionality $d$ of the lattice. When
this is not the case, from Eq. (\ref{eq:sigma_lowT}) a spontaneous
magnetization at a finite temperature should arise signaling the occurrence
of LRO. Of course, the low-$\mathbf{k}$ behaviors (\ref{eq:Omega_lowk})
determine the constraints for convergency of the integrals involved in the
low-$T$ expressions when $\overline{h}\rightarrow 0$. It is easy to see
that, in the limit $\overline{h}\rightarrow 0$, the integral (\ref
{eq:funcI(h)}) converges only for $d<p<2d$ with $d\leq 2$ and for $p>d$ with 
$d>2$. Then, for these values of $p$ and $d$, a spontaneous magnetization $%
m_{0}\left( T\right) =\sigma _{0}\left( T\right) S$ exists and hence LRO
occurs at small but finite temperature, with the result: 
\begin{equation}
\sigma _{0}\left( T\right) =\sigma \left( T,0\right) \simeq 1-\overline{T}%
I_{d}^{(p)}(0)
\end{equation}
where $I_{d}^{(p)}(0)$ is a finite quantity whose estimate can be
immediately obtained from Eq. (\ref{eq:sigma_explicit}). On the contrary,
for $\alpha \geq 2d$ with $d<2$, the integral (\ref{eq:funcI(h)}) diverges
as $\overline{h}\rightarrow 0$ and no finite solution for $\sigma $ exists
at $\overline{T}\neq 0$. This means that, for these values of $\alpha $ and $%
d$, no LRO occurs at finite temperature. The global situation is
schematically shown in Fig. 1. Here we also distinguished the domains where,
as we will show in the next Sec. 4, the system exhibits a critical behavior
like for a Heisenberg model with a nearest-neighbor exchange coupling
(short-range interaction (SRI) regime) and the long-range nature of
interactions (LRI) becomes effective.

In conclusion, our explicit low-temperature results suggest that a
transition to a FM phase at finite temperature occurs in the domains of the (%
$p-d$)-plane where LRO exists. In the remaining domains a different scenario
takes place with absence of a phase transition. These predictions are quite
consistent with the recent extension \cite{Bruno01} of the known
Mermin-Wagner theorem \cite{MerminWagner66} to quantum spin models with
LRI's of the type here considered.
Other thermodynamic properties can be also obtained within the framework of
CSDM when the solutions of the ME's are known. Indeed, for our spin model,
one can easily show that the internal energy $u$ and free energy $f$ per
spin are given by \cite{Cavallo02}: 
\begin{align}
u(T,h)& =\frac{\left\langle H\right\rangle }{N}  \nonumber \\
& =-\frac{h}{N}\left\langle S_{0}^{z}\right\rangle -\frac{1}{2N^{2}}\sum_{%
\mathbf{k}}J\left( \mathbf{k}\right) \left[ \left\langle S_{\mathbf{k}%
}^{+}S_{-\mathbf{k}}^{-}\right\rangle +\left\langle S_{\mathbf{k}}^{z}S_{-%
\mathbf{k}}^{z}\right\rangle \right] \text{,}  \label{eq:u} \\
f(T,h)& =f_{0}+\frac{1}{N}\int_{0}^{J}\frac{dJ^{\prime }}{J^{\prime }}%
\left\langle H_{I}\right\rangle \left( J^{\prime }\right)  \nonumber \\
& =f_{0}+\frac{1}{2N^{2}}\sum_{\mathbf{k}}\frac{J\left( \mathbf{k}\right) }{J%
}\int_{0}^{J}dJ^{\prime }\left[ \left\langle S_{\mathbf{k}}^{+}S_{-\mathbf{k}%
}^{-}\right\rangle _{J^{\prime }}+\left\langle S_{\mathbf{k}}^{z}S_{-\mathbf{%
k}}^{z}\right\rangle _{J^{\prime }}\right] \text{.}  \label{eq:f}
\end{align}
Here, $H_{I}$ is the interaction part of the spin Hamiltonian, $f_{0}$ is
the free energy per spin for a magnetic model without interactions and $%
\left\langle ...\right\rangle _{J^{\prime }}$ denotes a canonical average as
a function of the interaction strength $J^{\prime }$. As we see, within our
approximations all the quantities in Eqs. (\ref{eq:u}) and (\ref{eq:f}) can
be expressed in terms of the transverse SD $\Lambda _{\mathbf{k}}(\omega )$
and hence all the relevant thermodynamic quantities of our classical spin
model can be evaluated. In particular, for the reduced internal energy $%
\overline{u}=u/JS^{2}$ we have: 
\begin{equation}
\overline{u}=-\overline{h}\sigma -\frac{\overline{T}\sigma }{J}\int_{1BZ}%
\frac{d^{d}k}{(2\pi )^{d}}\frac{J\left( \mathbf{k}\right) }{\overline{\omega 
}_{\mathbf{k}}}-\frac{J(0)}{2J}\sigma ^{2}\text{.}  \label{eq:u2}
\end{equation}
Then, in the low temperature limit, from (\ref{eq:sigma_explicit}) we find: 
\begin{equation}
\overline{u}\simeq \overline{T}-\overline{h}-\frac{1}{2}\frac{J(0)}{J}%
+O\left( T^{2}g\left( \overline{h}\right) \right)  \label{eq:u2_lowk}
\end{equation}
where the explicit expression of the term $O\left( T^{2}g\left( \overline{h}%
\right) \right) $ can be obtained in a straightforward but tedious way.
Hence, for the reduced specific heat $\overline{C}_{h}=\left( \partial 
\overline{u}/\partial \overline{T}\right) _{h}=C_{h}/S$, as expected for a
classical spin model \cite{Cavallo02}, we have: 
\begin{equation}
\overline{C}_{h}\simeq 1+O\left( T^{2}g\left( \overline{h}\right) \right) 
\text{.}  \label{CC}
\end{equation}

\section{Near zero-magnetization regimes}
As we mentioned before, the decoupling $\left\langle S_{\mathbf{k}}^{z}S_{-%
\mathbf{k}}^{z}\right\rangle \approx \left\langle S_{\mathbf{k}%
}^{z}\right\rangle \left\langle S_{-\mathbf{k}}^{z}\right\rangle $ is
suitable for regimes with nonzero magnetization ($\sigma \neq 0$). Hence,
the basic equations used in the previous section do not allow us to explore
near-zero magnetization domains in the phase diagram as the critical region.
To overcome this difficulty, one is forced to find a more appropriate
decoupling procedure for the longitudinal correlation function which allows
us to obtain self-consistent ME's appropriate for describing regimes when $%
\sigma \rightarrow 0$ and preserves also the simplicity of the one $\delta $%
-function ansatz for the transverse SD $\Lambda _{\mathbf{k}}(\omega )$. A
possible and successful solution to this problem was suggested several years
ago for spin models with short-range interactions \cite
{Campana84,Kamienartz83}. These studies showed that a suitable decoupling
procedure when the magnetization approaches to zero (see also Ref. \cite
{Kalashnikov73} for the quantum counterpart) consists in writing (see Eq. (%
\ref{eq:ME2})): 
\begin{align}
& \frac{1}{N}\sum_{\mathbf{k}^{\prime }}\left[ J(\mathbf{k}^{\prime })-J(%
\mathbf{k}-\mathbf{k}^{\prime })\right] \left\langle S_{\mathbf{k}^{\prime
}}^{z}S_{-\mathbf{k}^{\prime }}^{z}\right\rangle \nonumber
\\
&  \simeq \frac{1}{N}\sum_{\mathbf{k}^{\prime }}\left[ J(\mathbf{k}^{\prime
})-J(\mathbf{k}-\mathbf{k}^{\prime })\right] \left\{ \left\langle S_{\mathbf{%
k}}^{z}\right\rangle \left\langle S_{-\mathbf{k}}^{z}\right\rangle -\frac{1}{%
2}\left( 1-\frac{\left\langle S_{0}^{z}\right\rangle ^{2}}{N^{2}S^{2}}%
\right) \left\langle S_{\mathbf{k}^{\prime }}^{+}S_{-\mathbf{k}^{\prime
}}^{-}\right\rangle \right\} \text{.}
\label{eq:decoupling}
\end{align}
As we see, with Eq. (\ref{eq:decoupling}) only the SD $\Lambda _{\mathbf{k}%
}(\omega )$ is involved and, inserting in the ME (\ref{eq:ME2}), one finds
for the reduced dispersion relation $\overline{\omega }_{\mathbf{k}}$ the
same formal expression (\ref{eq:se_R}) but with $R(\mathbf{k})$ replaced by: 
\begin{equation}
R(\mathbf{k})=1+\overline{T}\sigma \int_{1BZ}\frac{d^{d}k^{\prime }}{(2\pi
)^{d}}\frac{\left[ \Omega ^{\left( p\right) }\left( \mathbf{k}-\mathbf{k}%
^{\prime }\right) -\Omega ^{\left( p\right) }\left( \mathbf{k}^{\prime
}\right) \right] }{\overline{\omega }_{\mathbf{k}^{\prime }}\Omega ^{\left(
p\right) }\left( \mathbf{k}\right) }\text{.}  \label{eq:R(k)_new}
\end{equation}
Notice that, the effect of the decoupling (\ref{eq:decoupling}) corresponds
essentially to achieve in Eq. (\ref{eq:R(k)}) the transformation $1/(\sigma
N)\sum_{\mathbf{k}^{\prime }}\left( ...\right) \longrightarrow (\sigma
/N)\sum_{\mathbf{k}^{\prime }}\left( ...\right) $ \cite
{Campana84,Cavallo02,Kamienartz83}. Now, the new ME's can be properly used
for an estimate of the main critical properties of our classical spin model
when it exhibits LRO and of the low temperature paramagnetic susceptibility
in the remaining domains of the ($p-d$)-plane when no LRO exists.

\subsection{Critical temperature and critical behavior}
In the limit $\overline{h}\rightarrow 0$ with $\sigma \geq 0$, the system of
Eqs. (\ref{eq:se_R})-(\ref{eq:sigma}) with $R(\mathbf{k})$ given by Eq. (\ref
{eq:R(k)_new}), become 
\begin{align}
\overline{\omega }_{\mathbf{k}}& =\sigma \Omega ^{(p)}(\mathbf{k})R(\mathbf{k%
})\text{,}  \label{eq:omg2} \\
R(\mathbf{k})& =1+\frac{\overline{T}}{\Omega ^{(\alpha )}(\mathbf{k})}%
\int_{1BZ}\frac{d^{d}k^{\prime }}{(2\pi )^{d}}\frac{\Omega ^{(p)}(\mathbf{k}-%
\mathbf{k}^{\prime })-\Omega ^{(p)}(\mathbf{k}^{\prime })}{\Omega ^{(p)}(%
\mathbf{k}^{\prime })R(\mathbf{k}^{\prime })}\text{,}  \label{eq:R2} \\
\sigma ^{2}& =\frac{1-3\overline{T}Q(\overline{T})}{1-\overline{T}Q(%
\overline{T})}  \label{eq:sigma2}
\end{align}

where 
\begin{equation}
Q(\overline{T})=\int_{1BZ}\frac{d^{d}k}{(2\pi )^{d}}\frac{1}{\Omega ^{(p)}(%
\mathbf{k})R(\mathbf{k})}\text{.}  \label{eq:Q(T)}
\end{equation}
The reduced critical temperature $\overline{T}_{c}$ (with $T_{c}=JS^{2}%
\overline{T}_{c}$ ) can be determined by imposing the condition $\sigma (%
\overline{T}_{c})=0$ and hence by solving the self--consistent equation: 
\begin{equation}
1-3\overline{T}_{c}Q(\overline{T}_{c})=0\text{.}  \label{eq:find_Tc}
\end{equation}
For an explicit estimate of $\overline{T}_{c}$, one can again calculate the
integrals in Eqs. (\ref{eq:omg2})-(\ref{eq:Q(T)}) assuming for $\Omega
^{(p)}(\mathbf{k})$ the dominant contribution as $\mathbf{k}\rightarrow 0$
and solving our self-consistent equations by iteration. To first level of
iteration we find 
\begin{equation}
Q(\overline{T}_{c})\simeq \frac{I_{d}^{(p)}(0)}{1+\overline{T}%
_{c}I_{d}^{(p)}(0)}\text{.}  \label{eq:Q(Tc)}
\end{equation}
Then, Eq. (\ref{eq:find_Tc}) yields: 
\begin{equation}
\overline{T}_{c}=\frac{1}{2I_{d}^{(p)}(0)}
\end{equation}
where $I_{d}^{(p)}(0)$ is given by Eq. (\ref{eq:sigma_explicit}). Thus, a
critical temperature exists when $I_{d}^{(p)}(0)$ is finite and hence in the
domains of the ($p-d$)-plane where LRO takes place (see Fig. 1), as
expected. Of course the estimate for the critical temperature can be
systematically improved by calculating the integral $I_{d}^{(\alpha )}(0)$
using next $k$-powers in the expansions (\ref{eq:Omega_lowk}).

Here we consider explicitly the two-dimensional case for $2<p<4$ also for a
comparison with recent analytical and Monte Carlo results. In this case, the
coefficients in the expansions (\ref{eq:Omega_lowk}) are given by \cite
{Nakano95}:
\begin{equation}
\left\{ 
\begin{array}{l}
A_{2}=\frac{2^{2-p}\pi ^{2}}{\Gamma ^{2}(p)\sin \left[ \pi (p-2)/2\right] }
\\ 
B_{2}=2^{p-2}\zeta \left( \frac{p}{2}-1\right) \left[ \zeta \left( 1-\frac{p%
}{2},\frac{1}{4}\right) -\zeta \left( 1-\frac{p}{2},\frac{3}{4}\right) 
\right]
\end{array}
\right. \text{, \ }2<p<4
\end{equation}
where $\zeta (z,a)=\sum_{n=0}^{\infty }(n+a)^{-z}$ is the generalized
Riemann zeta function and $\zeta (z)\equiv \zeta (z,0)$ is the ordinary
Riemann zeta function. Then, for the integral $I_{2}^{(p)}(0)$, in the case
of interest $2<p<4$, we have: 
\begin{equation}
I_{2}^{(p)}(0)\simeq \frac{\Lambda ^{2-p}}{4-p}\frac{\Gamma ^{2}(p)}{%
2^{1-p}\pi ^{2}}\sin \left[ \frac{\pi (p-2)}{2}\right] ,
\end{equation}
and hence:
\begin{equation}
\overline{T}_{c}\left( d=2\right) =\frac{\left( 4-p\right) \pi ^{2}\Lambda
^{p-2}}{2^{p}\Gamma ^{2}(p)\sin \left[ \frac{\pi (p-2)}{2}\right] }.
\end{equation}
The critical temperature as a function of $p$ ($2<p<4$) for $d=2$ and $S=1/2$
is plotted in Fig. 2 and compared with the corresponding results recently
obtained for the quantum Heisenberg model by Nakano and Takahashi, using a
modified spin-wave (SW) theory \cite{Nakano94a} and the two-time Green
function EMM within the Tyablikov decoupling \cite{Nakano95}, and by
Vassiliev et al. with a Monte Carlo simulation \cite{Vassiliev01}. Our
result appears to be consistent with the ones obtained for the quantum
counterpart in view of the known feature that a quantum spin model can be
reasonably approximated by a classical one only in the large-$S$ limit \cite
{Stanley71}.

From Eqs. (\ref{eq:omg2})--(\ref{eq:Q(T)}) it is easy to obtain also the
behavior of $\sigma (\overline{T})$ as $T\rightarrow T_{c}^{+}$. With $%
m=\sigma S$ and $T=JS^{2}\overline{T}$, we get (for all values of $p$ and $d$
in the domains where a phase transition is allowed): 
\begin{equation}
m\sim \frac{(T-T_{c})^{\beta }}{T_{c}}  \label{m_beta}
\end{equation}
with 
\begin{equation}
\beta =\frac{1}{2}.
\end{equation}
Next, for the reduced paramagnetic susceptibility, defined as $\overline{%
\chi }=\sigma /\overline{h}=\chi /J$ in the limit $\overline{h}\rightarrow 0$
with $\sigma \rightarrow 0$, Eqs. (\ref{eq:se_R})-(\ref{eq:sigma}) with the
new expression (\ref{eq:R(k)_new}) for $R(\mathbf{k})$ reduce to 
\begin{align}
1& =3\overline{\chi }\overline{T}\int_{1BZ}\frac{d^{d}k}{(2\pi )^{d}}\frac{1%
}{1+\overline{\chi }X(\mathbf{k})}  \label{eq:X1} \\
X(\mathbf{k})& =\Omega ^{(p)}(\mathbf{k})+\overline{\chi }\overline{T}%
\int_{1BZ}\frac{d^{d}k^{\prime }}{(2\pi )^{d}}\frac{\Omega ^{(p)}(\mathbf{k}-%
\mathbf{k}^{\prime })-\Omega ^{(p)}(\mathbf{k}^{\prime })}{1+\overline{\chi }%
X(\mathbf{k}^{\prime })}  \label{eq:X2}
\end{align}
with $X(\mathbf{k})=\Omega ^{(p)}(\mathbf{k})R(\mathbf{k})$. As usual, an
estimate of $\overline{\chi }(\overline{T})$ as $\overline{T}\rightarrow 
\overline{T}_{c}^{+}$ can be obtained assuming the low-$\mathbf{k}$ behavior
(\ref{eq:Omega_lowk}) for $\Omega ^{(p)}(\mathbf{k})$ with the aim to find a
solution of Eqs. (\ref{eq:X1}) and (\ref{eq:X2}) such that $\overline{\chi }(%
\overline{T})\rightarrow \infty $ as $\overline{T}\rightarrow \overline{T}%
_{c}^{+}$. Then, we find: 
\begin{equation}
\chi =J\overline{\chi }\sim \left( \frac{T-T_{c}}{T_{c}}\right) ^{-\gamma }
\label{eq:chi-par1}
\end{equation}
with 
\begin{equation}
\gamma =\left\{ 
\begin{array}{ccc}
\frac{p-d}{2d-p},\text{ } & \frac{3}{2}d<p< & \left\{ 
\begin{array}{cc}
2d\text{,} & d\leq 2 \\ 
d+2\text{,} & d>2
\end{array}
\right. \\ 
1_{\ln }, & p=\frac{3}{2}d\text{ ,} & d\leq 4 \\ 
1, & d<p< & \left\{ 
\begin{array}{cc}
\frac{3}{2}d\text{,} & d<4 \\ 
d+2\text{,} & d>4
\end{array}
\right.
\end{array}
\right.
\end{equation}
and 
\begin{equation}
\gamma =\left\{ 
\begin{array}{ll}
\frac{2}{d-2},\text{ } & 2<d<4 \\ 
1_{\ln }, & d=4 \\ 
1, & d>4
\end{array}
,\right. p\geq d+2.
\end{equation}
The symbol $x_{\ln }$, here and below, denotes the main $(T-T_{c})$%
-dependence with a logarithmic correction (for instance, $\overline{\chi }%
\sim \left( \overline{T}-\overline{T}_{c}\right) ^{-1}\ln \left[ 1/\left( 
\overline{T}-\overline{T}_{c}\right) \right] $).

We now determine the behavior of the reduced magnetization $\sigma $ along
the critical isotherm as $\overline{h}\rightarrow 0$. Starting from the
basic ME's (\ref{eq:se_R})--(\ref{eq:sigma}) it is immediate to see that,
for small values of $\overline{h}$ and $\sigma $, the equation for the
critical isotherm has the form: 
\begin{equation}
1-\frac{2}{3}\sigma ^{2}-3\overline{T}_{c}Q_{c}(\sigma /\overline{h})=0
\label{eq:Qc1}
\end{equation}
where 
\begin{equation}
Q_{c}(\sigma /\overline{h})=(\sigma /\overline{h})\int_{1BZ}\frac{d^{d}k}{%
(2\pi )^{d}}\frac{1}{1+(\sigma /\overline{h})X_{c}(\mathbf{k})}
\label{eq:Qc2}
\end{equation}
and $X_{c}(\mathbf{k})$ is determined by Eq. (\ref{eq:X2}) with $\overline{%
\chi }$ replaced by $\sigma /\overline{h}$ and $\overline{T}=\overline{T}%
_{c} $. Eq. (\ref{eq:Qc1}) can be solved numerically but an reliable
estimate of the reduced magnetization $\sigma $ as $\overline{h}\rightarrow
0 $ can be simply obtained, as usual, assuming in Eqs. (\ref{eq:Qc1}) and (%
\ref{eq:Qc2}) the dominant contribution of $\Omega ^{(p)}(\mathbf{k})$ as $%
\mathbf{k}\rightarrow 0$. With this assumption and the definition $\sigma
\sim \overline{h}^{\frac{1}{\delta }}$ for the isotherm critical exponent $%
\delta $, Eqs. (\ref{eq:Qc1}) and (\ref{eq:Qc2}) yield: 
\begin{equation}
\delta =\left\{ 
\begin{array}{ccc}
\frac{p}{2d-p},\text{ } & \frac{3}{2}d<p< & \left\{ 
\begin{array}{cc}
2d\text{,} & d\leq 2 \\ 
d+2\text{,} & d>2
\end{array}
\right. \\ 
3_{\ln }, & p=\frac{3}{2}d\text{ ,} & d\leq 4 \\ 
3, & d<p< & \left\{ 
\begin{array}{cc}
\frac{3}{2}d\text{,} & d<4 \\ 
d+2\text{,} & d>4\text{,}
\end{array}
\right.
\end{array}
\right.
\end{equation}
and 
\begin{equation}
\delta =\left\{ 
\begin{array}{ll}
\frac{d+2}{d-2},\quad & 2<d<4 \\ 
3_{\ln },\quad & d=4 \\ 
3,\quad & d>4
\end{array}
,\right. p\geq d+2
\end{equation}
where $3_{\ln }$ characterizes the behaviour $\sigma \sim $ $\overline{h}%
^{1/3}\left| \ln \overline{h}\right| ^{1/3}$.

Finally, we calculate the critical exponent $\alpha $ for the specific heat
in the domains of $p$ and $d$ where the transition to a FM phase occurs. It
is important to note that we cannot use the expression (\ref{eq:u2}) for the
reduced internal energy per spin which has been obtained assuming the
decoupling $\left\langle S_{\mathbf{k}}^{z}S_{-\mathbf{k}}^{z}\right\rangle
\simeq \left\langle S_{\mathbf{k}}^{z}\right\rangle \left\langle S_{-\mathbf{%
k}}^{z}\right\rangle $. Rather, we must use the general relation (\ref{eq:u}%
) with the decoupling procedure (\ref{eq:decoupling}) appropriate near the
critical point. Bearing this in mind, the reduced internal energy per spin
near the critical point assumes the form: 
\begin{equation}
\overline{u}\simeq -\overline{h}\sigma -\frac{J(0)}{2J}\sigma ^{2}-\frac{1}{2%
}\overline{T}\sigma \left( 1+\sigma ^{2}\right) \int_{1BZ}\frac{d^{d}k}{%
(2\pi )^{d}}\frac{J\left( \mathbf{k}\right) /J}{\overline{\omega }_{\mathbf{k%
}}},  \label{eq:u_Ch}
\end{equation}
where now $\overline{\omega }_{\mathbf{k}}$ involves the quantity $R\left( 
\mathbf{k}\right) $ given by Eq. (\ref{eq:R(k)_new}). Working for $\overline{%
h}\rightarrow 0$, $\sigma \rightarrow 0$ with $\sigma /\overline{h}=%
\overline{\chi }\left( T\right) $ as $\overline{T}\rightarrow \overline{T}%
_{c}^{+}$, setting $\sigma \simeq \overline{\chi }\left( T\right) \overline{h%
}$ in Eq. (\ref{eq:u_Ch}) and taking into account Eq. (\ref{eq:X1}) for $%
\overline{\chi }\left( T\right) $ with $X\left( \mathbf{k}\right) $
evaluated as \textbf{k}$\rightarrow 0$, we have for the reduced zero-field
specific heat $\overline{C}_{\overline{h}=0}\left( \overline{T}\right)
=C_{h=0}\left( T\right) /S$: 
\begin{equation}
\overline{C}_{\overline{h}=0}\left( T\right) =\left( \frac{\partial 
\overline{u}}{\partial \overline{T}}\right) _{\overline{h}=0}\simeq \frac{1}{%
2}+\frac{1}{6}\overline{\chi }^{-2}\frac{\partial \overline{\chi }}{\partial 
\overline{T}},\text{ \ (}\overline{T}\rightarrow \overline{T}_{c}^{+}\text{).%
}  \label{eq:Ch0}
\end{equation}
This interesting expression allows us to determine the specific heat
critical exponent $\alpha $. Indeed, with $\overline{\chi }\left( \overline{T%
}\right) \simeq A\left( \overline{T}-\overline{T}_{c}\right) ^{-\gamma }$ or 
$\overline{\chi }\left( \overline{T}\right) \simeq A\left( \overline{T}-%
\overline{T}_{c}\right) ^{-1}\ln \left[ 1/(\overline{T}-\overline{T}_{c})%
\right] $, Eq. (\ref{eq:Ch0}) yields: 
\begin{equation}
\overline{C}_{\overline{h}=0}\left( T\right) \simeq \left\{ 
\begin{array}{ll}
\frac{1}{2}-\frac{1}{6}\gamma A^{2}\left( \overline{T}-\overline{T}%
_{c}\right) ^{-\alpha }, & \alpha =1-\gamma <0 \\ 
\frac{1}{2}-\frac{1}{6}A^{2}\left( \overline{T}-\overline{T}_{c}\right)
^{0}\ln ^{-1}\left[ \frac{1}{\overline{T}-\overline{T}_{c}}\right] ,\text{ }
& \alpha =0_{\ln }\text{ .}
\end{array}
\right.   \label{Ch_alpha}
\end{equation}
With these definitions and the values of the exponent $\gamma $ determined
before one can immediately obtain the desidered values of $\alpha $. We
find: 
\begin{equation}
\alpha =\left\{ 
\begin{array}{ccc}
\frac{3d-2p}{2d-p},\text{ } & \frac{3}{2}d<p< & \left\{ 
\begin{array}{cc}
2d\text{,} & d\leq 2 \\ 
d+2\text{,} & d>2
\end{array}
\right. \\ 
0_{\ln }, & p=\frac{3}{2}d\text{ ,} & d\leq 4 \\ 
0, & d<p< & \left\{ 
\begin{array}{cc}
\frac{3}{2}d\text{,} & d<4 \\ 
d+2\text{,} & d>4\text{,}
\end{array}
\right.
\end{array}
\right.
\end{equation}
in the domain where LRI's are active, and 
\begin{equation}
\alpha =\left\{ 
\begin{array}{ll}
\frac{d-4}{d-2},\quad & 2<d<4 \\ 
0_{\ln },\quad & d=4 \\ 
0,\quad & d>4
\end{array}
\right. ,\text{ }p\geq d+2
\end{equation}
in the SRI regime. One can easily check that the critical exponents $\beta $%
, $\alpha $, $\gamma $, and $\delta $ calculated above satisfy the well
known scaling laws $\gamma =\beta (\delta -1)$, $\alpha +2\beta +\gamma =2$
and $\alpha +\beta (1+\delta )=2$ for all values of $p$ and $d$ which allow
a transition to the FM phase. Besides, they coincide with those obtained by
Nakano and Takahashi \cite{Nakano95} for the quantum counterpart,
consistently with the universality hypothesis \cite{Stanley71}.

\subsection{Low--temperature paramagnetic susceptibility for $p\geq 2d$ and $%
d\leq 2$}
The low--temperature behavior of the susceptibility in absence of LRO, 
\textit{i.e.} for $p\geq 2d$ and $d\leq 2$ (see Fig. 1), is given by Eqs. (%
\ref{eq:X1}) and (\ref{eq:X2}) but bearing in mind that now $T_{c}=0$.
Following a procedure similar to the one used in the preceding section for
obtaining the behavior of the paramagnetic susceptibility as $T\rightarrow
T_{c}^{+}$, in the low--temperature limit, we obtain: 
\begin{equation}
\overline{\chi }\sim \left\{ 
\begin{array}{ll}
\exp \left[ \frac{A_{d}\Lambda ^{d}}{3\overline{T}}\right] ,\quad  & p=2d \\ 
\overline{T}^{-\frac{p-d}{p-2d}},\quad  & 2d<p<2+d \\ 
\overline{T}^{-\frac{2}{2-d}}\left( \ln \left[ \frac{1}{T}\right] \right) ^{%
\frac{d}{2-d}} & p=2+d \\ 
\overline{T}^{-\frac{2}{2-d}},\quad  & p>2+d.
\end{array}
\right.   \label{eq:Chi_Tdiv}
\end{equation}
The previous behaviors generalize those obtained in Ref. \cite{Cavallo02}
for $d=1$. It is worth noting that for case $p=2d$ one finds a pure
exponential divergence as $T\rightarrow 0$ as for one-dimensional model \cite
{Cavallo02}. This result, although consistent with that one obtained within
the modified spin-wave theory \cite{Nakano94a} and the EMM for the two-time
Green functions \cite{Nakano95} for the quantum counterpart, does not
reduce, for $d=1$, to the expression derived by Haldane \cite{Haldane88}
which contains a factor proportional to $\overline{T}^{-1/2}$. So, one must
expect a $d$-dependent power law factor in $\overline{T}$ in Eq. (\ref
{eq:Chi_Tdiv}), which corrects the pure low-temperature exponential
divergence. The simple lowest-order approximation in the CSDM, here used, is
not able to capture this important physical aspect. However, within the
spirit of the SDM \cite{Kalashnikov73}, one can hope to improve
systematically this result working to higher order approximations.

\section{Concluding remarks}
In this work we have applied the SDM, within the framework of the classical
statistical mechanics (CSDM), to study the thermodynamic properties of a
classical $d$-dimensional Heisenberg FM model with LRI's decaying as $r^{-p}$
($p>d$) in the presence of an external magnetic field. To lowest order in
the CSDM, the most relevant magnetic quantities have been obtained
analytically in the low-temperature regime under near saturation conditions
as a functions of $d$ and $p$. The FM LRO at finite temperature has been
shown to occurs in a wide region of the ($p-d$)-plane (see Fig. 1) where a
transition to a FM phase takes place decreasing the temperature.

The thermodynamic regimes with near zero and zero magnetization have been
explored on the basis of a proper modification in the ME's of the transverse
SD. So we have estimated the critical temperature and the main critical
properties of the model, beyond the mean field approximation, in the domains
of the ($p-d$)-plane where a FM ordered phase is expected. The critical
exponents, here determined as a function of $p$ and $d$, coincide with those
ones for the corresponding quantum model. This constitutes a microscopic
check of the irrelevance of quantum fluctuations, consistent with the
universality hypothesis in the theory of critical phenomena. In contrast, as
expected, our calculations show that the nonuniversal parameters, such as
the critical temperature, depends on the classical or quantum nature of the
model under study, especially for small values of the spin $S$ as explicitly
shown by a comparison of our critical temperature estimates with analytical
studies and Monte Carlo simulations previously achieved for the quantum
counterpart. Finally, we have determined the low-temperature behavior of the
paramagnetic susceptibility for $p\geq 2d$ with $d<2$ when no LRO occurs.

The rich scenario here obtained in a unified and consistent way for a
nontrivial spin model, shows clearly the potentiality and the effectiveness
of the SDM also in treating classical many-body systems. It is indeed
relevant feature that, already to the lowest-order approximation, the CSDM
is able to capture the essential physics of the model, consistent with Monte
Carlo simulations \cite{Vassiliev01,Romano89,Romano90} and exact results 
\cite{Kunz76,Frolich78,Rogers81,Simon81,Pfister81}, and to obtain results
beyond the Tyablikov-like approximation for arbitrary values of
dimensionality $d$ and the decaying exponent $p>d$. Besides, it offers the
possibility to achieve systematically higher-order approximations with the
aim to improve the results here obtained and to study also the damping of
the oscillations on the same footing \cite{Campana83}.

In conclusion, due to the great experience acquired in the quantum many-body
theory, we believe that the CSDM \cite
{Caramico81,Campana83,Campana84,Cavallo01,Cavallo02} and the formalism of
the two-time Green functions in classical statistical mechanics \cite
{Bogoljubov62} constitute a promising tool to explore equilibrium and
transport properties of a wide variety classical many-body (not only
magnetic) systems. Further developments and applications along this
directions are desiderable.

\newpage

\begin{figure}[tbp]
\begin{center}
\includegraphics*[width=9cm]{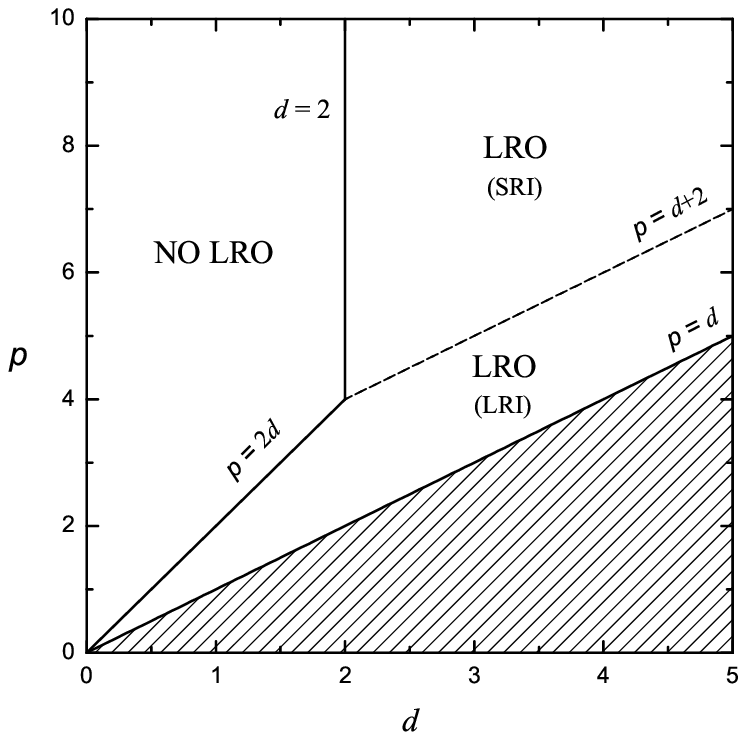}
\end{center}
\caption{Domains of the ($p-d$)-plane when a ferromagnetic long-range-order
(LRO) exists and is absent (NO LRO). The dashed line $p=d+2$ separates the
domains where long-range interaction (LRI)- and short-range interaction
(SRI)-regimes occur. The dashed region ($p<d$) corresponds to a nonextensive
thermodynamics. }
\label{fig:1_pdplane}
\end{figure}

\begin{figure}[tbp]
\begin{center}
\includegraphics*[width=10cm]{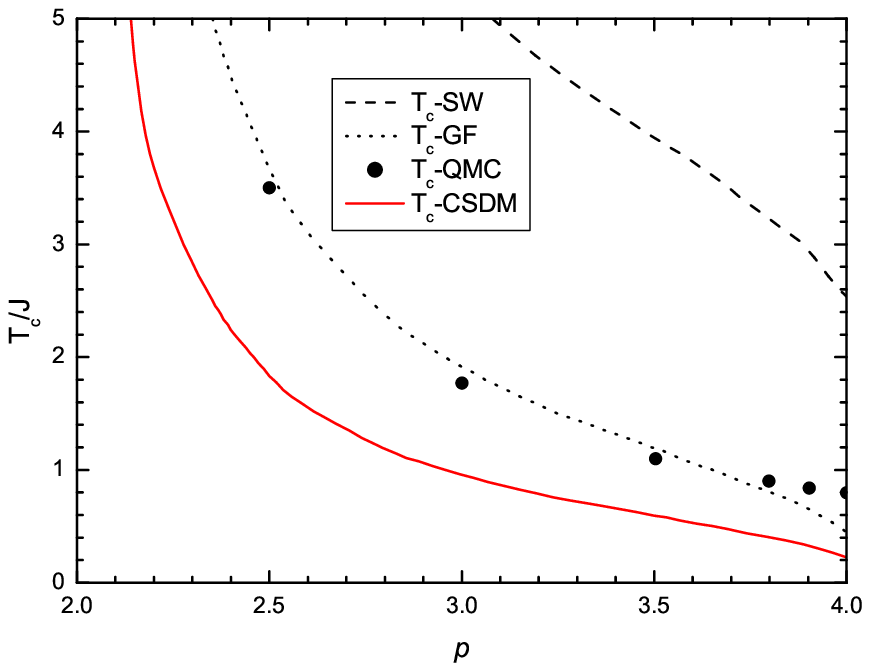}
\end{center}
\caption{The estimated critical temperature $T_{c}/J$ as a function of the
decaying exponent $p$ for a two-dimensional spin-$1/2$ Heisenberg
ferromagnet. The full line ($T_{c}-CSDM$) represents the estimate here
obtained. The dashed ($T_{c}-GF$) and ($T_{c}-SW$) lines refer to Green
function (GF) method \protect\cite{Nakano95} and modified spin-wave (SW)
theory \protect\cite{Nakano94a} results. The dots depict the quantum Monte
Carlo ($T_{c}-QMC$) predictions \protect\cite{Vassiliev01}. }
\label{fig:2_Tc}
\end{figure}


\begin{thebibliography}{00}
\bibitem{Zubarev60}  D.N. Zubarev, Usp. Fiz. Nauk. \textbf{71}, 71 (1960)
(Sov. Phys. Usp. \textbf{3}, 320 (1960))

\bibitem{Tyablikov67}  S.V. Tyablikov, \textit{Methods in the Quantum Theory
of Magnetism} (Plenum Pres, New York 1967); see also: N. Majlis, \textit{The
Quantum Theory of Magnetism} (World Scientific Singapore 2000).

\bibitem{Kalashnikov73}  O.K. Kalashnikov and E.S. Fradkin, Phys. Stat. Sol.
B \textbf{59}, 9 (1973) and references therein.

\bibitem{Bogoljubov62}  N.N. Bogoljubov and B.I. Sadovnikov, Zh. Eksp. Teor.
Fiz. \textbf{43}, 677 (1981) (Sov. Phys. JETP \textbf{16}, 482 (1963)).

\bibitem{Caramico81}  A. Caramico D'Auria, L. De Cesare and U. Esposito,
Phys. Lett. \textbf{85A}, 197 (1981).

\bibitem{Campana83}  L. S. Campana, A. Caramico D'Auria, M. D'Ambrosio, L.
De Cesare and U. Esposito, J. Phys. C \textbf{16}, L549 (1983).

\bibitem{Campana84}  L. S. Campana, A. Caramico D'Auria, M. D'Ambrosio, L.
De Cesare, G. Kamieniartz and U. Esposito, Phys. Rev. B \textbf{30}, 2769
(1984).

\bibitem{Cavallo01}  A. Cavallo, F. Cosenza, and L. De Cesare, Phys. Rev.
Lett. \textbf{87}, 240602 (2001); \textbf{88}, 099901(E) (2002).

\bibitem{Steiner76}  M. Steiner, J. Willain and C.G. Windsor, Adv. Phys. 
\textbf{25}, 87 (1976).

\bibitem{Reither80}  G. Reither and A. Sjolander, J. Phys. C \textbf{13},
32027 (1980).

\bibitem{Sepliarky01}  M. Sepliarky et al., Phys. Rev. B \textbf{64},
060101(R) (2001).

\bibitem{Haldane88}  F.D.M. Haldane, Phys. Rev. Lett. \textbf{60}, 635
(1988); Phys. Rev. Lett. \textbf{66}, 1529 (1991).

\bibitem{Shastry88}  B.S. Shastry, Phys. Rev. Lett. \textbf{60}, 639 (1988).

\bibitem{Nakano94a}  H. Nakano and M. Takahashi, Phys. Rev. B \textbf{50}
10331 (1994).

\bibitem{Nakano94b}  H. Nakano and M. Takahashi, J. Phys. Soc. Jpn. \textbf{%
63}, 4256 (1994).

\bibitem{Nakano95}  H. Nakano and M. Takahashi, Phys. Rev. B. \textbf{52},
6606 (1995).

\bibitem{Bruno01}  P. Bruno, Phys. Rev. Lett. \textbf{87}, 137203 (2001).

\bibitem{Vassiliev01}  O.N. Vassiliev, M.G. Cottam, and I.V. Rojdestvenski,
J. Appl. Phys. \textbf{89}, 7329 (2001) and references therein.

\bibitem{MerminWagner66}  N.D. Mermin and H. Wagner, Phys. Rev. Lett. 
\textbf{17}, 1133 (1966).

\bibitem{Kunz76}  H. Kunz and C.E. Pfister, Commun. Math. Phys. \textbf{46},
245 (1976).

\bibitem{Frolich78}  J. Fr\"{o}lich, R. Israel, H. Lieb, and B. Simon,
Commun. Math. Phys. \textbf{62}, 1 (1978).

\bibitem{Rogers81}  J.B. Rogers and C.J. Thompson, J. Stat. Phys. \textbf{25}%
, 669 (1981).

\bibitem{Simon81}  B. Simon, J. Stat. Phys. \textbf{26}, 307 (1981).

\bibitem{Pfister81}  C.E. Pfister, Commun. Math. Phys. \textbf{79}, 181
(1981).

\bibitem{Joyce66}  G.S. Joyce, Phys. Rev. \textbf{146}, 349 (1966).

\bibitem{Fisher72}  M.E. Fisher, S.-K. Ma and B.G. Nickel, Phys. Rev. Lett. 
\textbf{29}, 917 (1972).

\bibitem{Kosterlitz76}  J.M. Kosterlitz, Phys. Rev. Lett. \textbf{37}, 1577
(1976).

\bibitem{Romano89}  S. Romano, Phys. Rev. B \textbf{40}, 4970 (1989); 
\textit{ibid.} \textbf{46}, 5420 (1992).

\bibitem{Romano90}  S. Romano, Phys. Rev. B \textbf{42}, 6739 (1990); 
\textit{ibid.} \textbf{44}, 7066 (1991).

\bibitem{Cavallo02}  A. Cavallo, F. Cosenza, and L. De Cesare, Phys. Rev. B 
\textbf{66}, 174439 (2002).

\bibitem{Kamienartz83}  G. Kamienartz, Acta Phys. Pol. A \textbf{52}, 243
(1977); J. Phys. C \textbf{16}, 3763 (1983).

\bibitem{Stanley71}  H.E. Stanley, \textit{Introduction to Phase Transitions
and Critical Phenomena}, Clarendon Press, Oxford (1971).
\end{thebibliography}
\end{document}